# The Brightness of OneWeb Satellites

Anthony Mallama

anthony.mallama@gmail.com

2020 December 9

Abstract

The mean visual magnitude of OneWeb satellites at the standard satellite distance of 1,000 km is 7.18 +/-0.03 . When this value is adjusted to the nominal 1,200 km altitude of a OneWeb satellite in orbit it corresponds to magnitude 7.58 which is an indication of the mean brightness at zenith. The OneWeb satellites are fainter than the original Starlink satellites at a common distance. Preliminary data on the new and dimmer VisorSat design for Starlink suggests that they are still brighter than OneWeb at the satellites' respective operational altitudes.

1. Introduction

The many internet satellites in low-earth-orbit are a growing concern to astronomers because they can interfere with observation (Otarola et al., 2020, Walker et al., 2020, Tyson et al., 2020, Gallozzi et al., 2020, Hainaut and Williams, 2020 and McDowell, 2020). OneWeb currently has 74 satellites in space but the full constellation may eventually number more than a thousand. The SpaceX company is also launching a large internet fleet called Starlink and Amazon has announced similar plans.

The OneWeb satellite bus has a boxy shape with dish-like antennas mounted on short arms and a pair of large solar panels attached to longer arms. The altitude of their polar orbits is 1,200 km. By comparison the first Starlink satellites operate at just 550 km altitude. Thus a OneWeb satellite will be sunlit over a larger fraction of its orbit which can make it a greater concern to observers. On the other hand its greater height renders it less bright than a Starlink satellite as demonstrated in this paper.

Section 2 defines the magnitudes used herein and discusses the illumination phase angle (measured at the satellite between the Sun and the observer) which can effect brightness. Then Section 3 describes the observations along with the method of data processing. Section 4 characterizes the brightness of OneWeb satellites. Section 5 explains the limitations of this study. Section 6 compares the brightness of OneWeb satellites determined here with other measurements and compares the brightness of OneWeb with Starlink satellites.

This research is part of the author's collaboration with the Satellite Observations Working Group (Otarola et al., 2020). Most of the observations obtained and analyzed by the Working Group are acquired through spectral filters. The results of those studies are immediately useful to professional astronomers planning observations where interference by satellites is expected. This paper examines unfiltered magnitudes which measure brightness over all visible wavelengths. They may also be useful to professional astronomers in a general way, but they are more directly relevant to the public, especially amateur astronomers and naturalists.

2. Magnitudes and phase angles

The magnitudes in this paper are *visual.* That is, they represent the brightness of OneWeb satellites in combined red, green and blue colors with an effective wavelength in green light. Quantitatively, the band-pass sensitivity is centered near 0.6 micron and the width is about 0.4 micron. The V-band of the Johnson-Cousins photometric system is a close match to visual although the V bandwidth is narrower.

The *apparent magnitude* is that recorded by the sensor which, as in astronomy, has been corrected for atmospheric extinction. The satellite community also recognizes a measure of brightness called the *1000-km magnitude* which adjusts apparent brightness to that distance. Thus, it is analogous to the absolute magnitude used by astronomers.

Additionally, there is a *standard magnitude* which is meant to account for the illumination phase angle and gives the brightness of a half-illuminated satellite at 1000 km distance. The standard magnitude adjustment for phase angle is $2.5 * \log_{10} ( 0.5 + 0.5 \cos B )$. The term inside the parentheses is the fraction of a spherical object that is illuminated for an observer at phase angle B. In practice the standard magnitude is of limited value because few satellites are shaped like spheres. As noted above, OneWeb satellites are not spherical at all.

An empirical phase function may also be derived. In that case the relationship between the 1000-km magnitude and phase angle is determined by a least-squares fit to a mathematical expression. An empirical fit for OneWeb magnitudes is discussed in the next section.

Finally, since illumination geometry must contribute to satellite brightness other characterizations can be tried. For example, the Satellite Observations Working Group (Otarola et al., 2020) applies the Minnaert (1941) bidirectional reflectance function to observed magnitudes.

3. Observations and data processing

The Mini-MegaTORTORA (MMT) automated observatory (Karpov et al. 2015) is located in Russia at coordinates 44 degrees north and 41 degrees east. Each of the 9 MMT channels consists of a 71 mm diameter f/1.2 lens and a 2160 x 2560 sCMOS detector sensitive to visible light. A color transformation formula (Karpov, private communication) indicates that unfiltered MMT magnitudes are close to the Johnson-Cousins V band-pass for objects with small B-V color indices. Mallama (2020) found MMT magnitudes for Starlink satellites to be consistent with those of visual observers.

MMT posts satellite magnitudes and related information in an online database at http://mmt9.ru/satellites/. The author retrieved standard magnitudes and phase angles for all OneWeb satellites contained in the database as 2020 November 29.

The standard deviation of the 639 standard magnitudes was found to be 0.69. This indicates that the spherical-body phase function described in Section 2 does not accurately model the brightness of OneWeb satellites.

The author then computed 1000-km magnitudes by removing the spherical-body phase function correction from the standard magnitudes. Those results are plotted in Figure 1 along with linear and quadratic best fits which represent empirical phase functions. The root-mean-square scatter of the magnitudes without fitting is 0.68 which is just slightly better than that of the standard magnitudes. However, this only reduces to 0.67 for the linear fit and 0.66 for the quadratic fit. So, phase functions offer practically no improvement.

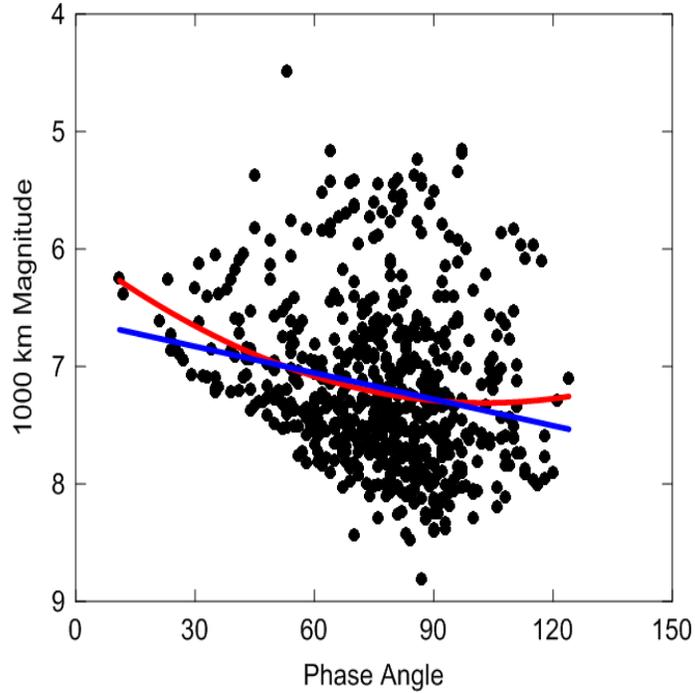

Figure 1. OneWeb 1000-km magnitudes plotted as a function of phase angle. Linear and quadratic fits are blue and red lines, respectively. The fits account for very little of the brightness variation.

4. Brightness characterization

The previous section demonstrated that neither the spherical-body nor the empirical phase functions account for any substantial portion of OneWeb satellite magnitude variation. Since those corrections and fits are of almost no value, the statistics of the simple 1000-km magnitudes are used to characterize brightness.

The mean and standard deviation of the 1000-km magnitudes are 7.18 and 0.68, respectively. The standard deviation of the mean is 0.03 magnitude. Adjusting to the nominal 1,200 km altitude of a OneWeb satellite in orbit gives magnitude 7.58 +/- 0.03 which is an indication of the mean brightness at zenith.

5. Limitations of this study

The standard magnitudes and phase angles used in the analysis are averages of those for each satellite track. The magnitudes were copied from the MMT database. The phase angles were estimated by eye from track plots; they are accurate to about 1 degree. A more rigorous (and time consuming) approach would be to associate every magnitude in every track with its corresponding phase angle.

The more rigorous approach might possibly reduce the scatter of processed magnitudes shown in Figure 1. However, experience with photometry of satellites having complex shapes suggests that the improvement would not be very great. The brightness of such satellites is difficult to model especially when their orientations are unknown.

Figure 2 illustrates that brightness can change erratically during a track. Here the apparent magnitudes have been converted to 1000-km and plotted versus phase angle. There are four large magnitude trends corresponding to brightness surges and two more smaller trends that cannot be explained by a phase angle correction. Large scale brightness variations are a common feature of OneWeb data.

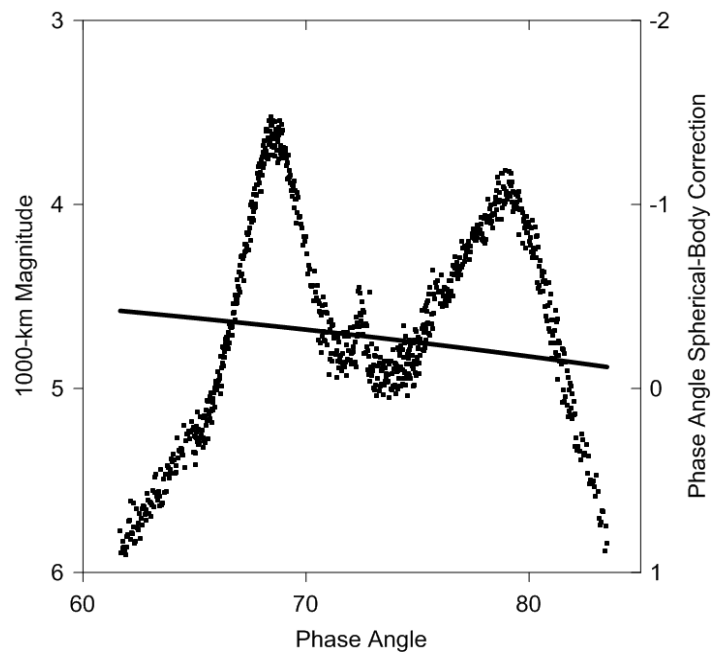

Figure 2. The 1000-km light-curve for a pass (circles) shown along with the phase angle correction for a spherical body (line). Both are plotted as a function of phase angle. The correction is too small to account for the observed magnitude variation.

A more thorough study could also try to make use of a detailed physical model that accounts for reflectance from each side of the satellite bus as well as its solar arrays and antennas. Cole (2020) formulated such a model for Starlink satellites. However the orientation of the satellite relative to the Sun and observer would also be required in order to predict its brightness.

6. Comparisons with other OneWeb magnitudes and with Starlink

There are two other sources of OneWeb magnitudes that may be compared with the results in this paper. The Satellite Observations Working Group (Otarola et al., 2020) reported a set of magnitudes but noted that "In most of the imaging attempts done at the POMENIS telescope, using a Johnson V spectral filter, the observations were too faint to obtain accurate photometry. The POMENIS team suspects the satellite streaks were fainter than around V=6 magnitude, which prevented detection by the automated POMENIS's software pipeline." Magnitudes fainter than 6 are consistent with the results reported in this paper.

The other source of data is an experienced visual observer who states that on a high elevation passes OneWeb satellites are magnitude 6-8 and that they are generally magnitude 7-8 (J. Respler, private communication). This range of brightness is also consistent with the results determined here.

The original Starlink satellites were very bright as reported by Mallama (2020) from an analysis of 830 visual magnitudes that included 554 measurements from MMT. The mean 1000-km magnitude was determined to be 5.93 +/-0.02 which is 1.25 magnitudes brighter than the equivalent OneWeb magnitude determined here. When that magnitude is adjusted to a distance of 550 km (the Starlink operational altitude) the result is 4.63 +/-0.02 which is 2.95 magnitudes brighter than that for OneWeb at its 1,200 km operational altitude.

SpaceX recently added a sun shade to their Starlink satellite design in order to reduce the brightness. The first of these new VisorSats have only been in orbit a short time and their brightness characteristics have not yet been carefully determined. However, early measurements indicate that they are between 1 and 2 magnitudes fainter than the original design. So, at 550 km they would still be brighter than OneWeb satellites at 1,200 km, while at the standard 1,000 km distance they would be approximately the same brightness.

7. Conclusions

This study examines the brightness of OneWeb satellites using magnitudes from the MMT observatory in Russia. There is considerable scatter (0.68 magnitude) in the calibrated data which is probably due to the complicated reflecting properties of this complexly shaped spacecraft. Nevertheless, the large number of observations allows for determination of a reliable mean magnitude of 7.18 +/-0.03 at the standard satellite distance of 1,000 km. When this value is adjusted to the nominal 1,200 km altitude of a OneWeb satellite in orbit it corresponds to magnitude 7.58 which is an indication of the mean brightness at zenith. OneWeb satellites are fainter than the original Starlink satellites. Preliminary results for the newer VisorSat design for Starlink indicate that they about the same brightness as OneWeb at a common distance. However, the lower altitude of VisorSats still renders them brighter in apparent magnitude than OneWeb satellites.

Acknowledgement

E. Katkova provided important information about the MMT system and data processing.

References

Cole, R.E. 2020. A sky brightness model for the Starlink "Visorsat" spacecraft. Research Note of the AAS. https://iopscience.iop.org/article/10.3847/2515-5172/abc0e9.

Gallozzi, S., Scardia, M., and Maris, M. 2020. Concerns about ground based astronomical observations: a step to safeguard the astronomical sky. https://arxiv.org/pdf/2001.10952.pdf.

Hainaut, O.R., and Williams, A.P. 2020. Impact of satellite constellations on astronomical observations with ESO telescopes in the visible and infrared domains. *Astron. Astrophys.* manuscript no. SatConst. https://arxiv.org/abs/2003.019pdf.

Karpov, S., Katkova, E., Beskin, G., Biryukov, A., Bondar, S., Davydov, E., Perkov, A. and Sasyuk, V. 2015. Massive photometry of low-altitude artificial satellites on minimegaTORTORA. Fourth Workshop on Robotic Autonomous Observatories. RevMexAA.

Mallama, A. 2020. Starlink satellite brightness before VisorSat. https://arxiv.org/abs/2006.08422.

Minnaert, M. 1941. The reciprocity principle in lunar photometry. The Astrophysical Journal, 93 403. 10.1086/144279.

McDowell, J. 2020. The low Earth orbit satellite population and impacts of the SpaceX Starlink constellation. *ApJ Let*, 892, L36 and https://arxiv.org/abs/2003.07446.

Otarola, A. (chairman) and Allen, L., Pearce, E., Krantz, H.R., Storrie-Lombardi, L., Tregloan-Reed, J, Unda-Sanzana, E., Walker, C. and Zamora, O. 2020. Draft Report of the Satellite Observations Working Group. Workshop on Dark and Quiet Skies for Science and Society commissioned by the United Nations, Spain and the International Astronomical Union. https://owncloud.iac.es/index.php/s/WcdR7Z8GeqfRWxG#pdfviewer.

Tregloan-Reed, J, Otarola, A., Ortiz, E., Molina, V., Anais, J., Gonzalez, R., Colque, J.P. and Unda-Sanza, E. 2020. First observations and magnitude measurement of SpaceX's Darksat. *Astron. Astrophys.*, manuscript no. Darksat_Letter_arXiv_submission_V2. https://arxiv.org/pdf/2003.07251.pdf.

Tyson, J.A., Ivezić, Ž., Bradshaw, A., Rawls, M.L., Xin, B., Yoachim, P., Parejko, J., Greene, J., Sholl, M., Abbott, T.M.C., and Polin, D. (2020). Mitigation of LEO Satellite Brightness and Trail Effects on the Rubin Observatory LSST. arXiv e-prints, arXiv:2006.12417.

Walker, C., Hall, J., Allen, L., Green, R., Seitzer, P., Tyson, T., Bauer, A., Krafton, K., Lowenthal, J., Parriott, J., Puxley, P., Abbott, T., Bakos, G., Barentine, J., Bassa, C., Blakeslee, J., Bradshaw, A., Cooke, J., Devost, D., Galadí-Enríquez, D., Haase, F., Hainaut, O., Heathcote, S., Jah, M., Krantz, H., Kucharski, D., McDowell, J.loan-Reed, J., Wainscoat, R., Williams, A., and Yoachim, P. (2020). Impact of Satellite Constellations on Optical Astronomy and Recommendations Toward Mitigations. Bulletin of the Astronomical Society, 52(2), 0206. 10.3847/25c2cfeb.346793b8.